\documentclass[conference]{IEEEtran}
\IEEEoverridecommandlockouts

\usepackage{cite}
\usepackage{amsmath,amssymb,amsfonts}
\usepackage{graphicx}
\usepackage{textcomp}
\usepackage{xcolor}
\usepackage{booktabs}
\usepackage{multirow}
\usepackage[hidelinks]{hyperref}
\usepackage{orcidlink}
\usepackage{tikz}
\usepackage{balance}
\usetikzlibrary{positioning, calc, arrows.meta}

\begin{document}
\bstctlcite{IEEEexample:BSTcontrol}
\title{A Pre-Dispatch Resonance Safety Criterion for AI Training Clusters \vspace{-0.5em}}

\author{%
\IEEEauthorblockN{%
Chandan~Chaudhary\orcidlink{0009-0002-2389-9568},
Abanish~Tiwari\orcidlink{0009-0003-0609-8571},~\emph{Student Member, IEEE},
Yansong~Pei\orcidlink{0000-0002-4647-7491},~\emph{Member, IEEE},\\
Mohammed~Ben-Idris\orcidlink{0000-0002-8731-8913},~\emph{Senior Member, IEEE},
and Joydeep~Mitra\orcidlink{0000-0001-9287-0983},~\emph{Fellow, IEEE}%
}
\IEEEauthorblockA{%
Electrical and Computer Engineering, Michigan State University, East Lansing, MI 48824, USA\\
E-mails: chaud152@msu.edu;\; tiwariab@msu.edu;\; peiyanso@msu.edu;\; benidris@msu.edu;\; mitraj@msu.edu%
}
\vspace{-3em}
}

\maketitle

\enlargethispage*{\baselineskip}
\begin{abstract}
  Hyperscale AI training clusters operate under the Bulk Synchronous Parallel protocol, which impose a periodic power swing on the transmission grid. Every GPU in the job transitions between compute and idle in lockstep, so the aggregate power traces a square wave at the training iteration period. Production iteration periods of one to ten seconds place the forcing frequency within the inter-area electromechanical mode band of large interconnections, where a training schedule can drive a mode at resonance. This paper derives a closed-form pre-dispatch safety criterion that bounds the maximum cluster size a grid can absorb at any proposed iteration period. The derivation inverts the steady-state forced two-area swing equations. The criterion defines a danger band of iteration periods, extends to the square-wave harmonics, and parameterizes the modal response from planning-study eigenanalysis and the forcing amplitude from GPU specifications. Applied to the IEEE 39-bus system at a production-representative duty cycle, the criterion shows that the maximum safe cluster at resonance is $66\,900$ GPUs under light damping. Rescheduling the same job less than one second away from resonance reduces the deviation $7.4\times$ with no hardware change. These results establish the training iteration period as a controllable grid-safety parameter and supply the analytic screening tool that reliability directives on current large loads lack.
% Applied to the IEEE 39-bus system at a production-representative duty cycle, the criterion shows that the maximum safe cluster at resonance is $66\,900$ GPUs under light damping ($\zeta_k = 0.03$), falling to $39\,300$ at the worst-case even duty. A $100\,000$-GPU cluster at resonance exceeds the $50$\,mHz absolute frequency threshold by $50\%$; the threshold is crossed within $9.8$\,s of job start, and the resonant tie-line swing reaches $266$\,MW from a $25$\,MW load fundamental. Rescheduling the same job less than one second away from resonance reduces the deviation $7.4\times$ with no hardware change. These results establish the training iteration period as a controllable grid-safety parameter and supply the analytic screening tool that reliability directives on computational loads lack.
\end{abstract}


\vspace{-0.75em}
\section{Introduction}
\label{sec:intro}
\enlargethispage{\baselineskip}
\enlargethispage{\baselineskip}
\enlargethispage{\baselineskip}

Power system dynamic security analysis has always treated load as a slowly varying, statistically smooth boundary condition. Inter-area electromechanical oscillations couple coherent groups of generators through energy exchange at 0.1--0.7Hz~\cite{kundur1994}. Dynamic security analysis accordingly treats them as a generation-side phenomenon, excited by faults or control interactions and contained by damping. Forced oscillations at these frequencies propagate interconnection-wide. A 200\,MW oscillation at 0.25\,Hz from a Florida steam turbine in 2019 reached New England at 50\,MW and persisted for 18 minutes~\cite{nerc2025largeloads}. Historically, such forced oscillations have originated from malfunctioning generation equipment. That assumption no longer holds.
 
The new forcing source is the AI training data center. Hyperscale training executes under the Bulk Synchronous Parallel (BSP) protocol~\cite{valiant1990bridging}. Every iteration decomposes into a compute phase, where each Graphics Processing Unit (GPU) runs near its thermal design power, and an AllReduce communication phase, where GPUs drop to near-idle power while gradients synchronize across all nodes~\cite{choukse2025power,go2025characterizing}. The AllReduce barrier forces every node to transition together~\cite{li2024unseen}. The aggregate power therefore traces a deterministic square wave at the training iteration period, whose reciprocal sets the forcing frequency. Production measurements on at-scale DGX-H100 jobs report swings that repeat from once per second to once every tens of seconds, spectral energy concentrated between 0.2 and 3\,Hz, and aggregate oscillations of tens of megawatts per datacenter~\cite{choukse2025power}. For H100 clusters of 100\,000 GPUs behind a 0.9-efficient power chain, the swing reaches 66.7\,MW at the point of common coupling~\cite{choukse2025power,go2025characterizing}. As the iteration period ranges from one to ten seconds, the forcing frequency sweeps 0.1--1\,Hz across the inter-area band. A North American Electric Reliability Corporation (NERC) Level~3 alert directs planners to identify vulnerable areas and operators to issue dispatch instructions to large computational loads~\cite{nerc2026level3alert,nerc2025level2alert}. No quantitative screen supports either obligation, nor can reactive detection substitute for one. A resonant job builds to the frequency threshold within seconds and sustains the violation for its full run duration~\cite{go2025characterizing}; the only controllable intervention point is before dispatch.
 
The existing literature surrounds this problem without solving it. A wide-area study models AI workload oscillations as stochastic periodic fluctuations, quantifies responses on the Western Electricity Coordinating Council (WECC) 179-bus system, and identifies fluctuation band and siting as severity factors~\cite{ko2026wide}. The framework predicts responses to assumed forcing but derives no pre-dispatch bound on cluster size. Operational risk assessments of data-center loads evaluate small-signal stability and parameter sensitivities~\cite{kwon2025lddl}. Their load profiles lack the single-tone BSP structure that drives resonance. The NERC large-loads whitepaper identifies these loads as sources of subsynchronous forced oscillations that are ``periodic, repetitive, and sustained,'' with reliability risks that are ``not always well defined''~\cite{nerc2025largeloads}. 
Our prior work has characterized the spatial correlation of AI data-center loads across transmission buses~\cite{chaudhary2026spatial}, identified their dominant dynamic bands using a Real-Time Digital Simulator (RTDS) testbed~\cite{chaudhary2026modal}, and quantified the resource adequacy risk associated with correlated large loads~\cite{chaudhary2026adequacy}. Three gaps motivate this work. First, no existing study treats the BSP job as a deterministic single-tone forcing source at the grid boundary whose frequency is set by the training schedule. Second, existing forced-oscillation studies typically start with a specified disturbance and evaluate the resulting system response. None inverts the analysis into a pre-dispatch bound on cluster size as a function of the proposed iteration period. Third, a criterion that distinguishes safe iteration periods from dangerous ones is needed to identify resonance-vulnerable areas and issue safe dispatch instructions. The literature offers none. 
\enlargethispage{\baselineskip}
\enlargethispage{\baselineskip}
\enlargethispage{\baselineskip}
\enlargethispage{\baselineskip}
 
This paper closes all three gaps with a pre-dispatch criterion derived by inverting the forced two-area swing equations. Three contributions follow.
 
\begin{enumerate}
  \item A grid-side BSP forcing model that reduces the square-wave power profile to a sinusoidal injection at the forcing frequency, grounded in production DGX-H100 measurements.
  \item A closed-form pre-dispatch criterion derived from the forced two-area swing equations, with harmonic-extended danger bands and threshold-crossing times.
  \item An IEEE 39-bus case study quantifying restriction severity, build-up times, and the recovery available through iteration-period rescheduling.
\end{enumerate}
 
The remainder of the paper is organized as follows. Section~\ref{sec:bspmodel} develops the BSP forcing model. Section~\ref{sec:criterion} derives the pre-dispatch criterion, danger band, and build-up time. Section~\ref{sec:casestudy} validates the criterion on the IEEE 39-bus system. Section~\ref{sec:conclusion} summarizes the findings and identifies directions for future work.
% \vspace{-0.25em}
\section{BSP Load Model}
\label{sec:bspmodel}
This section models the BSP power profile and derives the per-GPU forcing amplitude observed at the grid side.

\subsection{Iteration Structure and Power Profile}
\label{subsec:bsp_structure}
\begin{figure}[!htbp]
  \centering
  \vspace{-1em}
  \includegraphics[width=\columnwidth, trim={4cm 10.9cm 10.85cm 5.45cm},clip]{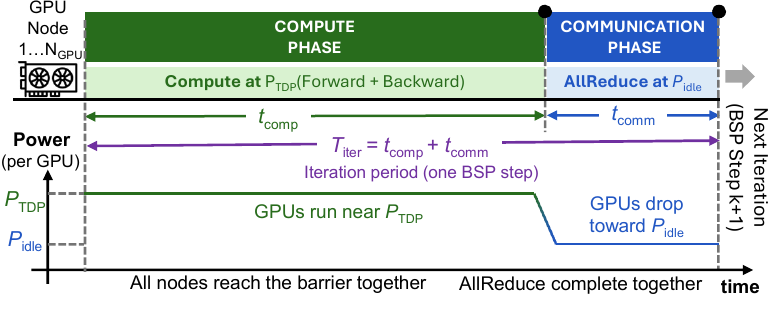}
  \vspace{-0.8cm}
  \caption{GPU power-phase behavior under one BSP training iteration.}
  \vspace{-0.5em}
  \label{fig:gpu}
\end{figure}

The BSP protocol partitions every training iteration into two phases executed identically by all GPU nodes in the job~\cite{valiant1990bridging}. The compute phase runs the forward and backward passes. All GPUs operate near thermal design power (TDP) $P_{\mathrm{TDP}}$ for a duration $t_{\mathrm{comp}}$. The communication phase runs the AllReduce collective, which aggregates gradients across all nodes. GPU power drops toward idle draw $P_{\mathrm{idle}}$ for a duration $t_{\mathrm{comm}}$~\cite{choukse2025power,go2025characterizing} (Fig.~\ref{fig:gpu}). The iteration period is $T_{\mathrm{iter}} = t_{\mathrm{comp}} + t_{\mathrm{comm}}$. Because the AllReduce barrier requires every node to complete its compute phase before any node advances, all $N_{\mathrm{GPU}}$ nodes transition between phases together~\cite{li2024unseen}. This synchronization is a protocol requirement enforced by the collective communication library on modern distributed training stacks~\cite{choukse2025power}.

\enlargethispage{\baselineskip}
\enlargethispage{\baselineskip}
\enlargethispage{\baselineskip}

The aggregate active power of the cluster thus traces a two-level periodic profile, which this work idealizes as the square wave
\begin{equation}
  P(t) = P_{\mathrm{floor}} + \Delta P \cdot s(t), \qquad s(t) \in \{0,1\},
  \label{eq:squarewave}
\end{equation}
where $P_{\mathrm{floor}} = N_{\mathrm{GPU}} P_{\mathrm{idle}} / \eta_{\mathrm{PS}}$ is the idle-phase IT load referred to the grid side, $\eta_{\mathrm{PS}}$ is the power supply chain efficiency from grid intake to GPU board, and the grid-side swing amplitude is
\begin{equation}
  \Delta P = \frac{N_{\mathrm{GPU}} \, (P_{\mathrm{TDP}} - P_{\mathrm{idle}})}{\eta_{\mathrm{PS}}}.
  \label{eq:deltaP}
\end{equation}
Conversion losses inflate the swing seen by the grid. The device-level difference is therefore divided, not multiplied, by $\eta_{\mathrm{PS}}$. For $\eta_{\mathrm{PS}} = 0.9$, representative of the UPS, distribution, and server power supply chain that ASHRAE documents~\cite{ashrae2016power}, each H100 contributes a 667\,W grid-side swing. The safe cluster bound scales inversely with $a$, so an $\eta_{\mathrm{PS}}$ anywhere in 0.85--0.95 shifts the bound by no more than 6\%.
The square wave $s(t)$ has duty cycle $d = t_{\mathrm{comp}} / T_{\mathrm{iter}}$, and its Fourier expansion places the fundamental at $f_c = 1/T_{\mathrm{iter}}$ with amplitude
\begin{equation}
  \Delta P_1 = \frac{2 \Delta P}{\pi} \sin(\pi d),
  \label{eq:fundamental}
\end{equation}
while harmonic $n$ carries amplitude proportional to $|\sin(n\pi d)|/n$. Inter-area modal responses concentrate at the fundamental because the harmonics carry less energy and usually fall above the inter-area band. Pre-dispatch screening therefore evaluates the sinusoidal injection $\Delta P_1 \sin(2\pi f_c t)$, with the harmonics incorporated through the duty-weighted harmonic extension of the criterion. The duty cycle is job-dependent. Dense configurations spend well over half the iteration in compute, while communication-bound mixture-of-experts configurations can exceed half in communication~\cite{go2025characterizing}. This study evaluates the representative dense-workload duty $d = 0.8$. The fundamental peaks at $d = 0.5$, where $\sin(\pi d) = 1$, so a communication-heavy job near even duty forces the grid hardest, and the criterion evaluated at $d = 0.5$ remains the absolute bound across all schedules.

Three systematic departures from~\eqref{eq:squarewave} occur in production, and each reduces the spectral line at $f_c$. Communication-phase power sits above true idle, runtime jitter spreads energy off the carrier tone, and power-smoothing rounds the phase transitions~\cite{choukse2025power,go2025characterizing,ko2026wide}. The ideal TDP-to-idle square wave therefore upper-bounds the fundamental amplitude of any production realization, so the criterion is conservative by construction. This is also the stress profile the industry applies. Power-stabilization studies emulate training with square-wave micro-benchmarks~\cite{choukse2025power}.

\subsection{Production Parameter Estimates}
\label{subsec:bsp_params}
\enlargethispage{\baselineskip}
\enlargethispage{\baselineskip}
\enlargethispage{\baselineskip}

Table~\ref{tab:bsp_params} reports representative parameters for H100 LLM training~\cite{choukse2025power,go2025characterizing}. Measured power swings repeat from once per second to once every tens of seconds, with spectral energy between 0.2 and 3\,Hz~\cite{choukse2025power}. Iteration periods of one to ten seconds place the fundamental in 0.1--1\,Hz, covering the inter-area band of large interconnections~\cite{kundur1994,rogers2000}.

\begin{table}[!htbp]
  \centering
  \vspace{-1em}
  \caption{H100 BSP load parameters~\cite{choukse2025power,go2025characterizing}}
  \label{tab:bsp_params}
  \vspace{-0.5em}
  \setlength{\tabcolsep}{4pt}
  \begin{tabular}{@{}llr@{}}
    \toprule
    \textbf{Parameter} & \textbf{Symbol} & \textbf{Value} \\
    \midrule
    GPU thermal design power      & $P_{\mathrm{TDP}}$   & 700\,W \\
    GPU idle power                & $P_{\mathrm{idle}}$  & $\approx$100\,W \\
    Power supply chain efficiency & $\eta_{\mathrm{PS}}$ & 0.90 \\
    Per-GPU swing (grid-side)     & $\Delta p$           & 667\,W \\
    Compute duty cycle (dense, this study) & $d$         & 0.80 \\
    Per-GPU fundamental ($d{=}0.8$) & $a$                & 249\,W \\
    % Per-GPU fundamental, worst case ($d{=}0.5$) & --     & 424\,W \\
    Iteration period range (inter-area) & $T_{\mathrm{iter}}$  & 1--10\,s \\
    Forcing frequency range (inter-area) & $f_c$              & 0.1--1\,Hz \\
    Cluster size (illustrative)   & $N_{\mathrm{GPU}}$   & 10\,000--100\,000 \\
    Square-wave swing (100k GPUs) & $\Delta P$           & 66.7\,MW \\
    Fundamental amplitude (100k, $d{=}0.8$) & $\Delta P_1$ & 24.9\,MW \\
    \bottomrule
  \end{tabular}
  \vspace{-1em}
\end{table}

\subsection{Grid-Side Representation}
\label{subsec:grid_side}
At the transmission level, the data center aggregates all GPUs, cooling loads, and power conversion stages into a single point of common coupling (PCC). Cooling dynamics occupy the band below 0.1\,Hz~\cite{chaudhary2026modal, chaudhary2025loadmodel} and converter-driven oscillations sit above 10\,Hz~\cite{mishra2025understanding}, both well outside the inter-area range. The grid therefore observes a single PCC perturbation $\delta P(t) = \Delta P_1 \sin(2\pi f_c t)$ riding on a slowly varying operating point, whose effect on system frequency the pre-dispatch criterion bounds. This representation isolates the BSP-driven IT oscillation. Non-IT loads, including HVAC, chillers, and auxiliaries, are included in the steady-state operating point, and their dynamics are not modeled.
\vspace{-0.25em}
\section{Pre-Dispatch Resonance Safety Criterion}
\label{sec:criterion}
\begin{figure}[!htbp]
  \centering
  \vspace{-1em}
  \begin{tikzpicture}[font=\scriptsize,
      box/.style={draw, rounded corners=1.5pt, align=center, inner sep=3pt},
      io/.style={box, fill=gray!12},
      term/.style={box, rounded corners=2.5mm, fill=gray!25},
      arr/.style={-{Stealth[length=2mm]}, semithick}]
    \node[io, text width=0.40\columnwidth] (job)
      {\textbf{Job specification}\\
       $N_{\mathrm{GPU}}$, $T_{\mathrm{iter}}$, $d$;\;
       $P_{\mathrm{TDP}}$, $P_{\mathrm{idle}}$, $\eta_{\mathrm{PS}}$};
    \node[io, text width=0.40\columnwidth, right=3mm of job] (grid)
      {\textbf{Planning-study modal data}\\
       $f_k$, $\zeta_k$, $E_1$, $E_2 \Rightarrow E_{\mathrm{eq}}$, $B$};
    \node[term, above=4mm of job] (start)
      {\textbf{Start: new training job request}};
    \node[box, text width=0.62\columnwidth, below=4mm of $(job.south)!0.5!(grid.south)$] (force)
      {\textbf{BSP forcing model}\\
       $\Delta P_1 = N_{\mathrm{GPU}}\, a(d)$ at $f_c = 1/T_{\mathrm{iter}}$};
    \node[box, text width=0.62\columnwidth, below=4mm of force] (crit)
      {\textbf{Criterion~\eqref{eq:Nstar}}\\
       evaluate $N^*(T_{\mathrm{iter}})$; \; safe if $N_{\mathrm{GPU}} \leq N^*$};
    \node[box, text width=0.36\columnwidth, below=5mm of crit, xshift=-15mm] (fix)
      {\textbf{Instruction}\\
       shift $T_{\mathrm{iter}}$ outside $\mathcal{D}$\\ or cap at $N^*$};
    \node[box, text width=0.24\columnwidth, below=5mm of crit, xshift=20mm] (go)
      {\textbf{Dispatch job}};
    \draw[arr] (start) -- (job);
    \draw[arr] (job.south) -- (job.south |- force.north);
    \draw[arr] (grid.south) -- (grid.south |- force.north);
    \draw[arr] (force) -- (crit);
    \draw[arr] (crit.south) -- ++(0,-2.5mm) -| (go.north)
      node[pos=0.5, above, fill=white, inner sep=1pt] {yes};
    \draw[arr] (crit.south) -- ++(0,-2.5mm) -| (fix.north)
      node[pos=0.5, above, fill=white, inner sep=1pt] {no};
    \draw[arr, dashed] (fix.west) -- ++(-10mm,0) |- (job.west)
      node[pos=0.25, sloped, above, inner sep=1.5pt] {revise job};
  \end{tikzpicture}
  \caption{Pre-dispatch screening workflow. }
  \label{fig:framework}
  \vspace{-1.25em}
\end{figure}
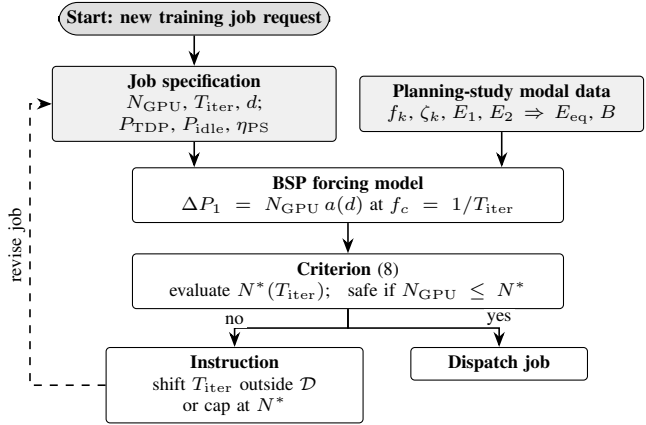

Fig.~\ref{fig:framework} summarizes the screening workflow. The criterion compares the planned cluster size against the safe bound and returns either a dispatch clearance or a schedule adjustment.

\enlargethispage{\baselineskip}
\enlargethispage{\baselineskip}
\enlargethispage{\baselineskip}

\subsection{Two-Area Swing Equations Under Periodic Forcing}
\label{subsec:swing}

The two-area model is the canonical representation of inter-area oscillations in interconnected power systems~\cite{kundur1994}. Let the grid consist of two coherent machine groups connected by a tie line. Their kinetic energies are $E_1 = H_1 S_1$ and $E_2 = H_2 S_2$, where $H_i$ is the inertia constant of group $i$ on its own MVA base $S_i$. The BSP cluster sits in group~1 and injects the sinusoidal perturbation $\delta P(t) = \Delta P_1 \sin(2\pi f_c t)$. The linearized swing equations for the relative angle $\delta(t) = \delta_1(t) - \delta_2(t)$ reduce to a single forced second-order oscillator~\cite{kundur1994,rogers2000},
\begin{equation}
  \ddot{\delta} + 2\zeta_k \omega_k \dot{\delta} + \omega_k^2 \delta
  = \frac{\omega_s \, B}{2 E_{\mathrm{eq}}} \, \Delta P_1 \sin(2\pi f_c t),
  \label{eq:swing_forced}
\end{equation}
where $\omega_k = 2\pi f_k$ is the inter-area modal angular frequency, $\zeta_k$ is the modal damping ratio, $\omega_s = 2\pi f_s$ is the synchronous angular frequency with $f_s = 60$\,Hz, $B = E_2 / (E_1 + E_2)$ is the participation of group~1 in the mode, and $E_{\mathrm{eq}} = E_1 E_2 / (E_1 + E_2)$ is the equivalent kinetic energy. The steady-state angle amplitude at $f_c$ follows from the standard forced-oscillator solution. The relative frequency $\dot{\delta}/(2\pi)$ between the two areas carries one factor of $B$ from the forcing term in~\eqref{eq:swing_forced}. The absolute frequency deviation at group-1 buses follows from the center-of-inertia decomposition. Specifically, $\Delta f_1 = [E_2/(E_1+E_2)]\,\dot{\delta}/(2\pi) = B\,\dot{\delta}/(2\pi)$, which introduces a second factor of $B$. The absolute frequency deviation at group-1 buses therefore has amplitude
\begin{equation}
  |\Delta f_1(f_c)| = \frac{B^2 \, \Delta P_1 \, f_s \, f_c}
  {4\pi E_{\mathrm{eq}} \sqrt{(f_k^2 - f_c^2)^2 + (2\zeta_k f_k f_c)^2}},
  \label{eq:freq_dev}
\end{equation}
where $\Delta f_{\max}$ in~\eqref{eq:tcross} and~\eqref{eq:Nstar} refers to the absolute area-1 frequency deviation compared against NERC BAL-001-2. At exact resonance ($f_c = f_k$) the radical collapses to $2\zeta_k f_k^2$ and the amplitude becomes
\begin{equation}
  |\Delta f_1|_{\mathrm{res}} = \frac{B^2 \, \Delta P_1 \, f_s}{8\pi E_{\mathrm{eq}} \, \zeta_k f_k},
  \label{eq:freq_dev_res}
\end{equation}
which exceeds the response to the same injection at $f_c \ll f_k$ by the quality factor $Q_k = 1/(2\zeta_k)$. The same oscillation appears on the tie line through the synchronizing coefficient $T_s = 2 E_{\mathrm{eq}} \omega_k^2 / \omega_s$. The tie-line power swing is $\Delta P_{\mathrm{tie}} = T_s |\delta|$, which at resonance reduces to $\Delta P_{\mathrm{tie}} = Q_k B \, \Delta P_1$. The grid-side amplification is therefore visible directly in megawatts. From rest, the resonant response grows with the envelope $|\Delta f|_{\mathrm{res}}\,(1 - e^{-\zeta_k \omega_k t})$. The time to reach a threshold $\Delta f_{\max} < |\Delta f|_{\mathrm{res}}$ is
\begin{equation}
  t_{\times} = \frac{1}{\zeta_k \omega_k} \ln\!\left(\frac{|\Delta f|_{\mathrm{res}}}{|\Delta f|_{\mathrm{res}} - \Delta f_{\max}}\right).
  \label{eq:tcross}
\end{equation}

\enlargethispage{\baselineskip}
\enlargethispage{\baselineskip}
% \enlargethispage{\baselineskip}

\subsection{Closed-Form Safety Criterion}
\label{subsec:criterion_derivation}
A pre-dispatch criterion requires $|\Delta f(f_c)| \leq \Delta f_{\max}$, where $\Delta f_{\max}$ is an operator-specified deviation threshold. Substituting the fundamental $\Delta P_1 = N_{\mathrm{GPU}} \, a$ with the duty-explicit per-GPU amplitude $a = (2/\pi)(P_{\mathrm{TDP}} - P_{\mathrm{idle}})\sin(\pi d)/\eta_{\mathrm{PS}}$ into~\eqref{eq:freq_dev} and solving for $N_{\mathrm{GPU}}$ yields the criterion
\begin{equation}
  N^*(T_{\mathrm{iter}}) = \frac{4\pi E_{\mathrm{eq}} \, \Delta f_{\max}
  \sqrt{(f_k^2 - f_c^2)^2 + (2\zeta_k f_k f_c)^2}}
  {B^2 \, f_s \, f_c \, a},
  \label{eq:Nstar}
\end{equation}
where $f_c = 1/T_{\mathrm{iter}}$. The $B^2$ denominator reflects both the forcing projection ($B$ in the RHS of~\eqref{eq:swing_forced}) and the area-1 frequency decomposition ($B$ in $\Delta f_1 = B\,\dot{\delta}/(2\pi)$). The criterion depends solely on modal system parameters ($f_k$, $\zeta_k$, $E_{\mathrm{eq}}$, $B$), device-level GPU parameters that enter through $a$, and the proposed scheduling parameter $T_{\mathrm{iter}}$. All inputs are available from standard planning analyses and the job specification. An operator evaluates~\eqref{eq:Nstar} before dispatching a training job. The job is grid-safe if $N_{\mathrm{GPU}} \leq N^*(T_{\mathrm{iter}})$. Otherwise the operator shifts $T_{\mathrm{iter}}$ away from $1/f_k$ or reduces the cluster size. At resonance the criterion reaches its minimum,
\begin{equation}
  N^*_{\min} = \frac{8\pi E_{\mathrm{eq}} \, \Delta f_{\max} \, \zeta_k f_k}{B^2 \, f_s \, a}.
  \label{eq:Nstar_min}
\end{equation}
Because $N^*_{\min}$ scales as $\zeta_k f_k E_{\mathrm{eq}}$, the binding constraint comes from lightly damped, low-frequency modes in low-inertia conditions. A mode with $\zeta_k = 0.03$ restricts the cluster $3.3\times$ harder than the same mode at $\zeta_k = 0.10$.

\subsection{Danger Band and Off-Resonance Recovery}
\label{subsec:danger_band}

The danger band in iteration-period space is the set of $T_{\mathrm{iter}}$ values for which the planned cluster size exceeds the criterion,
\begin{equation}
  \mathcal{D} = \{T_{\mathrm{iter}} : N^*(1/T_{\mathrm{iter}}) < N_{\mathrm{GPU}}^{\mathrm{plan}}\}.
  \label{eq:danger_band}
\end{equation}
Because $N^*(f_c)$ has a single minimum at $f_c = f_k$, the danger band is a contiguous interval centered near $T_k = 1/f_k$. Its width grows with $N_{\mathrm{GPU}}^{\mathrm{plan}}$ and with $1/\zeta_k$. Lower damping deepens the resonance notch in $N^*(T_{\mathrm{iter}})$, so a fixed planned cluster size violates the criterion over a wider interval of iteration periods. Outside $\mathcal{D}$ the criterion is satisfied at the planned size, so moving $T_{\mathrm{iter}}$ out of the band restores unrestricted operation without any hardware change. $T_{\mathrm{iter}}$ is a free scheduling variable, adjusted through gradient accumulation or batch sizing in the training launch configuration. %Avoiding the danger band costs only that adjustment.

\enlargethispage{\baselineskip}
\enlargethispage{\baselineskip}

\subsection{Assumptions and Validity Range}
\label{subsec:assumptions}

Criterion~\eqref{eq:Nstar} rests on three assumptions. First, the two-area model captures one dominant inter-area mode. Systems with several modes require~\eqref{eq:Nstar} per mode, with the binding constraint taken as the minimum over all $k$. The square wave also forces the mode through its harmonics. Harmonic $n$ carries the per-GPU amplitude $a_n = (2/\pi)(P_{\mathrm{TDP}} - P_{\mathrm{idle}})\,|\sin(n\pi d)|/(n\,\eta_{\mathrm{PS}})$ at frequency $n f_c$, so the complete screen evaluates $N^*_n(T_{\mathrm{iter}}) = [\,n \sin(\pi d)/|\sin(n\pi d)|\,] \, N^*$ at $n f_c$ and takes the minimum over $n$. At $d = 0.5$ only odd harmonics exist and $N^*_n = n N^*$. Off even duty the even harmonics return, and at $d = 0.8$ the second harmonic carries 0.81 of the fundamental amplitude while the fifth vanishes because $\sin(4\pi) = 0$. Section~\ref{sec:casestudy} applies this harmonic-extended form. Second, the steady-state response dominates after the transient decays with time constant $1/(\zeta_k \omega_k)$. For $f_k = 0.6$\,Hz and $\zeta_k \in [0.03, 0.10]$, three time constants span 8--27\,s, a small fraction of any training run, and BSP iteration periods hold stable over the run~\cite{go2025characterizing}. Third, the swing dynamics remain in the linear region. At the threshold the relative angle amplitude is $|\delta| = \Delta f_{\max}/(B f_c) = 0.050/(0.639 \times 0.60) = 0.130$\,rad at $f_c = f_k = 0.6$\,Hz, well inside the linear range. The criterion therefore enforces its own validity.
% \vspace{-0.5em}
\section{Case Study and Discussion}
\label{sec:casestudy}

This section applies the criterion to the IEEE 39-bus system and discusses the implications for operators and planners.
\enlargethispage{\baselineskip}
\enlargethispage{\baselineskip}
\enlargethispage{\baselineskip}

\subsection{Test System and Two-Area Parameters}
\label{subsec:test_system}
The IEEE New England 39-bus system~\cite{athay1979practical,pai1989energy} contains ten generators. The IEEE benchmark validation of this system reports nine electromechanical modes and classifies exactly one as inter-area: eight modes are local or regional within the New England area, and only the lowest-frequency mode, near 0.6\,Hz, crosses the area boundary, with generators 2--10 oscillating coherently against generator~1, the New York equivalent~\cite{canizares2016benchmark}. The system is therefore a physical two-area structure, and the criterion of Section~\ref{sec:criterion} applies without modal truncation. The data center connects at bus~12~\cite{chaudhary2026spatial,chaudhary2026modal}. Bus~12 carries negligible conventional load in the standard 39-bus model, so the data center is the dominant injection at that bus. 

\begin{table}[!htbp]
  \centering
  \vspace{-1.25em}
  \caption{Two-area parameters for the IEEE 39-bus case study}
  \label{tab:modal_params}
  \setlength{\tabcolsep}{5pt}
  \begin{tabular}{@{}llr@{}}
    \toprule
    \textbf{Parameter} & \textbf{Symbol} & \textbf{Value} \\
    \midrule
    Inter-area modal frequency & $f_k$ & 0.60\,Hz \\
    Resonance iteration period & $T_k = 1/f_k$ & 1.67\,s \\
    Modal damping ratio (swept) & $\zeta_k$ & 0.03, 0.05, 0.10 \\
    Quality factor & $Q_k = 1/(2\zeta_k)$ & 16.7, 10.0, 5.0 \\
    Kinetic energy, generators 2--10 & $E_1$ & 28.27\,GW$\cdot$s \\
    Kinetic energy, generator 1 & $E_2$ & 50.0\,GW$\cdot$s \\
    Equivalent kinetic energy & $E_{\mathrm{eq}}$ & 18.06\,GW$\cdot$s \\
    Group-1 participation & $B$ & 0.639 \\
    Deviation threshold & $\Delta f_{\max}$ & 50\,mHz \\
    \bottomrule
  \end{tabular}
  \vspace{-1em}
\end{table}

At 100\,000 GPUs the facility draws a peak grid-side IT load of 77.8\,MW, holds an idle-phase floor of 11.1\,MW, and swings 66.7\,MW between phases. This swing is 1.1\% of the 6.1\,GW system load, too small to register in steady-state screening; its grid significance lies in its switching frequency. Generator~1 at bus~39 is not a physical plant but the aggregated New York interconnection, which is why its inertia is an order of magnitude above every other unit in the model~\cite{canizares2016benchmark,athay1979practical}. The two-area parameters in Table~\ref{tab:modal_params} follow from the published machine inertias~\cite{athay1979practical,pai1989energy}; a BSP cluster within the New England group sees $E_{\mathrm{eq}} = 18.06$\,GW$\cdot$s and $B = 0.639$. Modal damping varies with dispatch, load level, and stabilizer tuning, so the study sweeps $\zeta_k \in \{0.03, 0.05, 0.10\}$, spanning the lightly damped to well-damped inter-area behavior reported in interconnection oscillation measurements~\cite{nerc2019oscillation,kundur1994}.

\subsection{Resonance Map: \texorpdfstring{$N^*(T_{\mathrm{iter}})$}{N*(T\_iter)}}
\label{subsec:resonance_map}
\begin{figure}[!htbp]
  \centering
  \vspace{-1em}
  \includegraphics[width=\columnwidth]{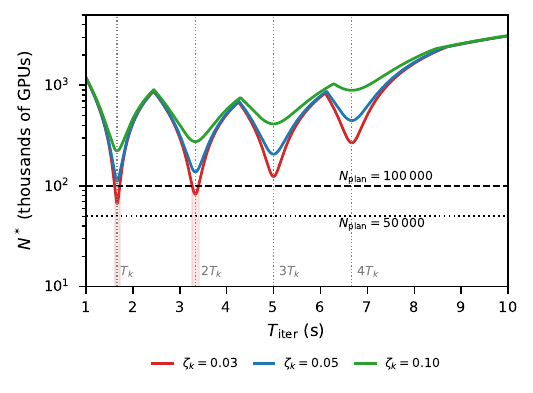}
  \vspace{-1cm}
  \caption{Pre-dispatch criterion for the inter-area mode ($d = 0.8$); secondary notches at $nT_k$ arise from the square-wave harmonics.}
  \vspace{-1em}
  \label{fig:resonance_map}
\end{figure}

Fig.~\ref{fig:resonance_map} plots the criterion~\eqref{eq:Nstar} over $T_{\mathrm{iter}} \in [1, 10]$\,s for the three damping ratios, with the per-GPU fundamental $a = 249$\,W at the study duty $d = 0.8$ from Table~\ref{tab:bsp_params} and $\Delta f_{\max} = 50$\,mHz, just inside the Eastern Interconnection Frequency Trigger Limit of $3\varepsilon_1 = 54$\,mHz that bounds Balancing Authority ACE limits under NERC Standard BAL-001-2~\cite{nercbal0012}. Each curve forms a notch at $T_k = 1.67$\,s whose depth grows as damping falls. At resonance, the criterion bounds the safe cluster to $N^*_{\min} = 66\,900$ GPUs for $\zeta_k = 0.03$, $111\,500$ for $\zeta_k = 0.05$, and $222\,900$ for $\zeta_k = 0.10$; a communication-heavy job at $d = 0.5$ tightens every floor by the factor $1/\sin(0.8\pi) = 1.70$, to $39\,300$ GPUs at light damping. The two lighter-damping floors sit below the scale of current frontier training deployments. The three curves merge away from the notch because damping enters the response only through the $2\zeta_k f_k f_c$ term, which the detuning term $(f_k^2 - f_c^2)^2$ dominates off resonance. The restriction releases quickly: at $T_{\mathrm{iter}} = 2.0$\,s the fundamental criterion allows $423\,500$ GPUs at $\zeta_k = 0.05$. Release is not monotone, because the harmonics of the square wave re-excite the mode, and at $d = 0.8$ the even harmonics participate. The harmonic-extended screen forms secondary notches at $T_{\mathrm{iter}} = 2T_k = 3.3$\,s, $3T_k = 5.0$\,s, and $4T_k = 6.7$\,s with floors $1.24\,N^*_{\min}$, $1.85\,N^*_{\min}$, and $4\,N^*_{\min}$; the fifth harmonic vanishes at this duty. At $\zeta_k = 0.03$ the second harmonic places the $100\,000$-GPU cluster in a secondary band over $T_{\mathrm{iter}} \in [3.27, 3.40]$\,s; at $\zeta_k \geq 0.05$ the harmonic floors exceed $100\,000$\,GPUs and no secondary band exists. The resonance risk for a $100\,000$-GPU cluster at $\zeta_k = 0.03$ therefore occupies two narrow windows: $[1.61, 1.72]$\,s near the fundamental and $[3.27, 3.40]$\,s near the second harmonic, both inside the production scheduling range reported in~\cite{choukse2025power}.

\enlargethispage{\baselineskip}
\enlargethispage{\baselineskip}
\enlargethispage{\baselineskip}

\subsection{Danger Bands and Restriction Severity}
\label{subsec:danger_bands}

Table~\ref{tab:danger_bands} reports the danger bands computed from~\eqref{eq:danger_band} for a planned cluster of $N_{\mathrm{GPU}}^{\mathrm{plan}} = 100\,000$. At $\zeta_k = 0.03$ the floor of $66\,900$ GPUs lies below the planned size, producing a fundamental danger band of $0.11$\,s; the floor rises to $111\,500$ at $\zeta_k = 0.05$ and $222\,900$ at $\zeta_k = 0.10$, so no fundamental band exists for those damping levels at $100\,000$ GPUs. The restriction tightens strictly with cluster scale: a 66\,900-GPU cluster sits exactly at the $\zeta_k = 0.03$ floor, while a 50\,000-GPU cluster clears all three damping levels at this duty. Resonant severity follows the quality factor. At $100\,000$ GPUs the absolute frequency deviation at area-1 buses reaches $74.8$\,mHz at $\zeta_k = 0.03$, $44.9$\,mHz at $\zeta_k = 0.05$, and $22.4$\,mHz at $\zeta_k = 0.10$. Only the lightly damped case violates the 50\,mHz threshold. At $d = 0.5$ worst duty the floors tighten by $1.70\times$, and both $\zeta_k = 0.03$ and $\zeta_k = 0.05$ then exceed the threshold at $100\,000$ GPUs.

\begin{table}[!htbp]
  \centering
    \vspace{-2em}
  \caption{Resonance severity and danger bands}
  \label{tab:danger_bands}
  \setlength{\tabcolsep}{4pt}
  \begin{tabular}{@{}cccccc@{}}
    \toprule
    $\zeta_k$ & $Q_k$ & $N^*_{\min}$ & $|\Delta f_1|_{\mathrm{res}}$ (mHz) & $\mathcal{D}$ (s) & Width (s) \\
    \midrule
    0.03 & 16.7 & 66\,900 & 74.8 & [1.61, 1.72] & 0.11 \\
    0.05 & 10.0 & 111\,500 & 44.9 & --           & --   \\
    0.10 &  5.0 & 222\,900 & 22.4 & --           & --   \\
    \bottomrule
  \end{tabular}
  \vspace{-1.5em}
\end{table}
\enlargethispage{\baselineskip}
\enlargethispage{\baselineskip}
% \enlargethispage{\baselineskip}

\subsection{Resonant vs.\ Off-Resonance Scheduling}
\label{subsec:time_domain}
\begin{figure}[!htbp]
  \centering
  \vspace{-1em}
  \includegraphics[width=\columnwidth]{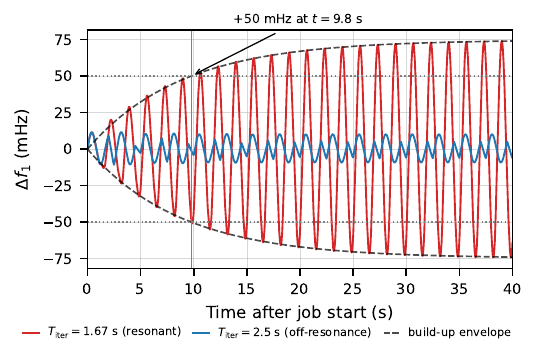}
  \vspace{-2em}
  \caption{Absolute frequency deviation at area-1 buses for $N_{\mathrm{GPU}} = 100\,000$, $\zeta_k = 0.03$, $d = 0.8$: resonant ($T_{\mathrm{iter}} = T_k$) vs.\ off-resonance ($T_{\mathrm{iter}} = 2.5$\,s).}
  \vspace{-0.5em}
  \label{fig:time_domain}
\end{figure}

Fig.~\ref{fig:time_domain} shows the numerically integrated response for the $100\,000$-GPU cluster at $\zeta_k = 0.03$, $d = 0.8$, under a resonant schedule ($T_{\mathrm{iter}} = 1.67$\,s) and an off-resonance one ($T_{\mathrm{iter}} = 2.5$\,s). The resonant response grows along the envelope of~\eqref{eq:tcross} with time constant $1/(\zeta_k \omega_k) = 8.8$\,s and settles at $74.8$\,mHz, $1.5\times$ the 50\,mHz threshold. The second and third harmonics contribute $2.4$\,mHz and $0.9$\,mHz at 1.2 and 1.8\,Hz; the numerically integrated square-wave steady-state of $74.4$\,mHz agrees with the analytic fundamental to within $0.5$\%, which validates the screening model of Section~\ref{sec:bspmodel} at resonance. The oscillation crosses the threshold $t_{\times} = 9.8$\,s after job start, within six iterations.

At $\zeta_k = 0.05$ and $\zeta_k = 0.10$ no crossing occurs at this cluster size and duty, consistent with the absence of danger bands in Table~\ref{tab:danger_bands}. On the tie line the resonant swing is $\Delta P_{\mathrm{tie}} = Q_k B \Delta P_1 = 266$\,MW at $\zeta_k = 0.03$ and $159$\,MW at $\zeta_k = 0.05$, sustained oscillations comparable to and above the 200\,MW source amplitude of the 2019 Florida event~\cite{nerc2025largeloads}. At $T_{\mathrm{iter}} = 2.5$\,s the fundamental falls to $5.4$\,mHz and the tie-line swing to $29$\,MW; the second harmonic lands at 0.8\,Hz and contributes an additional $6.2$\,mHz, so the simulation peak settles at $10.1$\,mHz, a $7.4\times$ reduction from a 0.83\,s rescheduling of the iteration period. The simulation peak (10.1\,mHz) lies below the arithmetic sum of the two contributions (11.6\,mHz) because the fundamental and second-harmonic components do not reach their individual peaks simultaneously. Off-resonance placement must clear the mode for every significant harmonic, which is what the harmonic-extended screen checks. The same cluster, the same hardware, and the same site move from violation to compliance through one scheduling parameter.

\enlargethispage{\baselineskip}
\enlargethispage{\baselineskip}

\subsection{Discussion}
\label{subsec:discussion}
The defining feature of these resonance violations is their persistence. A training job holds its iteration period for the full run duration~\cite{go2025characterizing}, so a resonant schedule sustains its forced oscillation for hours rather than for the seconds of a fault. This class of disturbance does not appear in load-flow or transient stability screens. It is neither a contingency nor an instability, but the steady-state response to a load that planning studies represent as constant. As data center capacity grows, this gap between what planning screens and what is actually injected will widen. The invisibility also rules out reactive control. Post-event detection faces a measurement problem~\cite{chatterjee2026pnnl}. By the time a resonant tone is confirmed, the violation has already been sustained for minutes. Every input the criterion needs is available before energization from modal analysis and the job specification. The danger band translates those inputs into a concrete operator instruction (schedule outside $\mathcal{D}$, e.g., outside $[1.61, 1.72]$\,s at $\zeta_k = 0.03$ for a $100\,000$-GPU cluster, or cap the cluster at $N^*$) that satisfies NERC Essential Actions \#2 and \#7 and is verifiable from telemetry~\cite{nerc2026level3alert}.

The criterion also clarifies which mitigation is worth pursuing and at what cost. Hardware remedies such as battery buffering and GPU power smoothing can suppress the oscillation amplitude at the PCC but carry capital cost or efficiency loss~\cite{choukse2025power}. Iteration-period adjustment through gradient accumulation steps or batch size is a configuration change with no hardware cost, though it shifts the compute-to-communication balance and may affect training throughput. The criterion makes the tradeoff explicit. Outside the danger band no intervention is needed. Inside it, the operator can present both options to the load entity with quantified bounds on each, rather than defaulting to expensive hardware as a precaution. This changes the conversation between grid operators and data center operators from a vague reliability concern to a specific, negotiable scheduling constraint.
\begin{figure}[!htbp]
  \centering
  \vspace{-1.5em}
  \includegraphics[width=0.9\columnwidth]{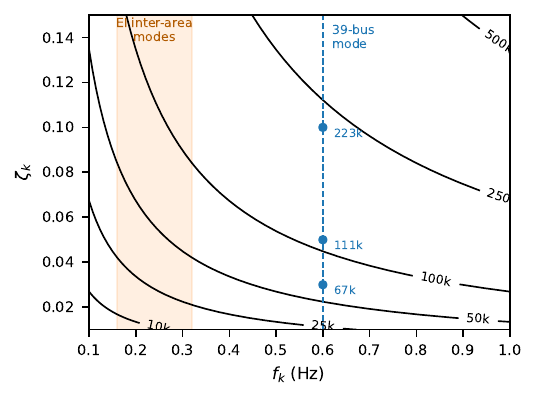}
  \vspace{-1.65em}
  \caption{Resonant-floor contours $N^*_{\min}$ (thousands of GPUs) over the $(f_k, \zeta_k)$ plane at the 39-bus kinetic energy.}
  \label{fig:floor_chart}
  \vspace{-1em}
\end{figure}

The floor chart (Fig.~\ref{fig:floor_chart}) shows that the risk is not specific to the 39-bus system and that its severity depends on where the grid is headed, not just where it is. The resonant floor scales linearly with modal damping, modal frequency, and system inertia, so the contours shift as grid conditions change. On this test system the danger band sits below 2\,s, at the fast edge of production iteration periods, so most jobs at this scale would not be affected. The measured inter-area modes of the Eastern Interconnection lie at 0.16--0.32\,Hz~\cite{nerc2019oscillation}, placing resonance periods at 3.1--6.3\,s, which is the middle of the production scheduling window. Modes in that band impose floors roughly three times tighter than the 0.6\,Hz mode studied here, so at interconnection scale resonance screening is not a niche check but a routine one. The trend also matters. As inverter-based resources displace synchronous machines and system inertia falls, the safe cluster size shrinks in proportion. The same training job that clears the criterion today may bind it during a low-inertia dispatch hour, and the criterion makes that future restriction visible in advance.
\vspace{-0.25em}
% \enlargethispage{\baselineskip}
% \enlargethispage{\baselineskip}
% \enlargethispage{\baselineskip}
\section{Conclusion}
\label{sec:conclusion}

This paper establishes that the resonance risk a synchronized training cluster poses to the grid is a property of the schedule, and that it is computable before the job runs. The forcing frequency is set by a configuration parameter, so the same cluster is dangerous or benign depending on a number in its launch script. The derived criterion determines whether the job is safe using quantities available before energization. Screening therefore requires no new measurement infrastructure, no load model beyond the fundamental of the synchronized power swing, and no simulation. A single algebraic inequality separates safe from resonance-dangerous operating points. On the benchmark system the inequality binds at cluster sizes already in service, and it identifies the schedule shift that restores compliance at no hardware cost. The training iteration period thus becomes a controllable scheduling quantity alongside redispatch and curtailment, and the criterion tells both the operator and the load owner when to reach for it.

The criterion treats one dominant inter-area mode at a fixed operating point, bounds the load by its worst-case envelope, and absorbs all non-IT dynamics into the operating point. Systems with closely spaced modes, dispatch-dependent modal parameters, drifting duty cycles, and coupled facility loads define the limits of its validity. Future work will validate the criterion against time-domain simulations and facility data, develop danger-band tracking under shifting modal parameters, and extend the method to interconnection-scale systems. Pre-dispatch screening of training schedules belongs alongside every other study the grid performs before accepting a new source of forcing.

% \section*{Acknowledgment}
% This work was supported in part by the MSU Research Foundation.
% \vspace{-1em}
% \newpage
% \balance
% \enlargethispage{\baselineskip}
\enlargethispage{\baselineskip}

% \begingroup
% \let\oldthebibliography\thebibliography
% \renewcommand{\thebibliography}[1]{%
%   \oldthebibliography{#1}%
%   \vspace{-0.19em}%
%   % \enlargethispage{\baselineskip}
% }
\bibliographystyle{IEEEtran}
\bibliography{references}
% \endgroup
\end{document}